\documentclass[12pt]{article}
\usepackage{blindtext}
\usepackage{amsmath}
\usepackage{amssymb}
\usepackage{wrapfig}
\usepackage{multicol}
\usepackage[margin=1in]{geometry}
\usepackage{graphicx}
\usepackage{caption}
\usepackage{indentfirst}
\usepackage{slashed}

\graphicspath{{FIGS/}}
\emergencystretch\maxdimen
\renewenvironment{abstract}
{\par\noindent\textbf{\abstractname:}\ \ignorespaces}
{\par\medskip}

\makeatletter
\renewcommand\maketitle
{\noindent
{\large\bfseries\@title}%
\medskip\par
{\@author}%
\medskip\par
{\@date}%
\bigskip\par\noindent
}

\renewcommand\section{\@startsection {section}{1}{\z@}%
{-1.5ex}%
{0.05ex }%
{\bfseries}}
\makeatother

\begin{document}
\onecolumn
\title{A momentum space formulation for some relativistic statistical field theories with quantum-like observables}
\author{Brenden McDearmon }
\maketitle

\begin{abstract}
Considering a  fluctuating scalar field on momentum space, some relativistic statistical field theories are constructed. A Hilbert space of observables is then constructed from functionals of the fluctuating scalar field with an inner product defined in terms of expectation values of the functionals. A bosonic Fock space is then constructed from the Hilbert space and creation and annihilation operators that act on the Fock space are defined. The creation and annihilation operators are used to define field operators. These field operators have some interesting quantum-like properties. For example, the field operators do not commute in general and, in the particular case of the free field theory, can be shown to satisfy the microcausality condition. 
\end{abstract}

\section*{Introduction}

Quantum field theory has been studied from various perspectives. By one perspective, a classical relativistic field theory is ``quantized'' by replacing the classical field with corresponding field operators that act on some Hilbert space with imposed commutation or anti-commutation relations that depend on whether the field is fermionic or bosonic \cite{peskin2018introduction}\cite{itzykson2012quantum}. In this canonical quantization picture, observables are calculated in terms of expectation values of the field operators with respect to some state in the Hilbert space \cite{peskin2018introduction}\cite{itzykson2012quantum}. For example, the LSZ reduction formula relates expectation values of time ordered products of field operators to scattering amplitudes \cite{peskin2018introduction}\cite{itzykson2012quantum}.

By a second perspective, these same scattering amplitudes can in turn be calculated using the Feynman path integral  \cite{peskin2018introduction}\cite{itzykson2012quantum}\cite{bailin1993introduction}. In the Feynman path integral approach, the expectation values of time ordered products of field operators are calculated by averaging functionals of the classical field over all possible field configurations weighted by a complex factor $e^{iS^m \slash \hbar}$ involving the matter action $S^m$ \cite{peskin2018introduction}\cite{itzykson2012quantum}\cite{bailin1993introduction}. While some techniques exist for evaluating these real time Feynman path integrals, the oscillatory nature of the complex-valued path integral measure can make numerical evaluation of these expectation values challenging \cite{berges2007lattice} \cite{berges2021complex}. To make these calculations more tractable, the real time Feynman path integrals are often ``Wick rotated'' to imaginary time to provide a real-valued path integral measure \cite{bailin1993introduction}.

Wick rotation converts the real time Feynman path integral on Minkowski spacetime into an imaginary time path integral on 4-dimensional Euclidean space \cite{roepstorff2012path}  \cite{glimm2012quantum}. This provides another perspective on quantum field theory whereby expectation values of observables are calculated by averaging the observable over all configurations of the Euclidean classical field \cite{roepstorff2012path}  \cite{glimm2012quantum}. This averaging is performed using a Euclidean path integral having a real-valued probability density proportional to $e^{- \beta S_E^m}$ where $S_E^m$ is the Wick rotated Euclidean matter action and $\beta$ is a real parameter analogous to the thermodynamic $\beta$ (i.e., inverse temperature) of classical statistical mechanics \cite{roepstorff2012path}  \cite{glimm2012quantum}. An advantage of this Euclidean approach to quantum field theory is that it is readily amenable to non-perturbative calculations by discretizing the system with a Euclidean lattice and evaluating the Euclidean path integral numerically \cite{montvay1994quantum}. A disadvantage, however, is that Euclidean space lacks the causal structure of Minkowski spacetime. Despite this disadvantage,  it is sometimes possible to reconstruct relativistic observables from the Euclidean expectation values \cite{glimm2012quantum} \cite{osterwalder1973axioms} \cite{osterwalder1975axioms}. 

It would be desirable to have a formulation of quantum field theory that is both manifestly relativistic, as in the Feynman path integral approach, and has a real-valued probability density, as in the Euclidean path integral approach. Indeed, some authors have suggested that such a reformulation of quantum field theory may be possible. See, e.g., \cite{tilloy2017interacting}, \cite{floerchinger2010quantum}, \cite{hitschfeld2019probability}, and \cite{kent2013path}. If such a perspective of quantum field theory as a relativistic statistical field theory is possible, it may be more amenable to non-perturbative calculations, owing to the real-valued probability density. Indeed the recent article by Giachello and Gradenigo \cite{giachello2024symplectic} provides some numerical calculations for a relativistic field theory using their ``symplectic quantization'' scheme. See, also, \cite{gradenigo2024symplectic2}, \cite{gradenigo2021symplectic1}, and \cite{gradenigo2021symplectic2}. This ``symplectic quantization'' is similar to the ``variational dynamics'' algorithm used in \cite{mcdearmon2023euclidean1} and \cite{mcdearmon2023euclidean2}, the ``variational dynamics'' additionally including an ``action bath'' coupled to the system. Extending this ``variational dynamics'' approach to a relativistic system, this article constructs a relativistic statistical field theory on momentum space and demonstrates that the resulting algebra of observables has some interesting quantum-like properties\footnote{In a companion article, a spacetime formulation is presented \cite{mcdearmon2024rel}.}.

\section*{Constructing a relativistic statistical field theory on momentum space}

Let $\mathcal{\tilde{M}}$ be a 4-dimensional manifold with points $p := \left(p_0, p_1, p_2, p_3\right)$, and equip $\mathcal{\tilde{M}}$ with the Minkowski metric $p \cdot p :=  p_0^2 -p_1^2- p_2^2 -p_3^2$ . A relativistic statistical field theory will be defined for a matter field, $\phi$, having support on a 3-dimensional subset $V$ of $\mathcal{\tilde{M}}$. The 3-dimensional subset $V$ can be defined as a subset of the zero set of some relativistically invariant function, $P(p, \phi)$, which may, for nonlinear theories, also depend on $\phi$. A particular example discussed in more detail below is that of a real-valued scalar field $\phi$ defined on the positive mass-shell of the Klein-Gordon equation such that $V=V_m^{free} := \{ p \in \mathcal{\tilde{M}} \, | \, p \cdot p -m^2=0  \text{ with } p_0 \geq 0\}$ where $P(p, \phi)=P^{free}_m(p):=p \cdot p -m^2$.

Suppose that the field $\phi$ is allowed to ``fluctuate'' with respect to some parameter $\lambda \in [0, \infty)$. That is,  $\phi(p, \lambda) \in \mathbb{R}$ for all $ (p, \lambda) \in \mathcal{\tilde{M}} \times [0, \infty)$. To prescribe how $\phi$ fluctuates with respect to $\lambda$, it is helpful to consider the derivative $\pi^{(\phi)}(p, \lambda) :=\left.\frac{d \phi(p, \lambda+\epsilon)}{d\epsilon}\right|_{\epsilon=0}$. 

Suppose now that these two fields, $\phi$ and $\pi^{(\phi)}$, exchange action between each other and an ``action bath'' as they fluctuate. Let the exchange of action between the fields and the action bath be mediated by fluctuations of a global real and positive scalar, $s(\lambda) \in (0, \infty)$, and its derivative $\pi^{s}(\lambda) :=\left.\frac{d s(\lambda+\epsilon)}{d\epsilon}\right|_{\epsilon=0}$. This collection of degrees of freedom defines a variational phase space $\Gamma := \{\times_{p \in V} \,  \phi(p),\times_{p \in V} \, \pi^{(\phi)}(p), s, \pi^{(s)}\}$. 

An action for this system is a real-valued function on the variational phase space, $S\left(\gamma \left(\lambda\right)\right) \in  \mathbb{R}$ for all $\gamma(\lambda) \in \Gamma$. Let the action be of the form $S := s(\lambda) \left( S^x(\lambda)-S^0 \right)$ where $S^0$ is the action of the system evaluated at $\lambda =0$ and $S^x$ is defined by the following equation:

\small
 \begin{equation}
\begin{aligned}
 S^x(\lambda) := \int_{V} \frac{\left(\pi^{(\phi)}(p, \lambda)\right)^2}{ 2\left(s(\lambda)\right)^2} d^3p + \frac{\left(\pi^{(s)}(\lambda)\right)^2}{2m_s} + S^m[\phi]+ \frac{n_f}{\beta} ln\left(s(\lambda)\right).
\end{aligned}
\end{equation}
\normalsize

\noindent Here, $n_f$ is a number of degrees of freedom of the matter field $\phi$, $m_s$ is a coupling constant, $0<\beta<\infty$ is a parameter that is analogous to the thermodynamic $\beta$ (i.e., inverse temperature) of classical statistical mechanics, and $S^m[\phi]$ is a matter action. For example, a free matter action can be defined on $V_m^{free} $ using a Gaussian matter action given by $S^m_{free}[\phi]:=  \int_{V^{free}_m} \frac{1}{2} \left(\phi(p)\right)^2 d^3p$. 

The action, $S$, can be treated mathematically as a Hamiltonian function on the variational phase space with the Hamiltonian differential equations (i.e., equations (2)-(5) below) providing a symplectic flow on the variational phase space that conserves the total action of the system. See, e.g., \cite{mcdearmon2023euclidean1} and \cite{mcdearmon2023euclidean2}.

\small
 \begin{equation}
\begin{aligned} 
\dot{\pi}^{(\phi)}(p, \lambda)  :=& \left.\frac{d \pi^{(\phi)}(p, \lambda+ \epsilon)}{d \epsilon}\right|_{\epsilon=0} = -\left.\frac{\delta S}{\delta\phi}\right|_{\lambda} 
\end{aligned}
 \end{equation}

 \begin{equation} 
\begin{aligned}
\dot{\phi}(p, \lambda) := \left.\frac{d \phi(p, \lambda+ \epsilon)}{d \epsilon} \right|_{\epsilon=0} = \left.\frac{\delta S}{\delta \pi^{(\phi)}}\right|_{\lambda} = \frac{\pi^{(\phi)}(p, \lambda)}{s(\lambda)},
\end{aligned}
\end{equation}

 \begin{equation} 
\begin{aligned}
\dot{\pi}^{(s)}(\lambda) := \left.\frac{d \pi^{(s)}(\lambda+ \epsilon)}{d \epsilon} \right|_{\epsilon=0} = -\left.\frac{\partial S}{\partial s}\right|_{\lambda} =   \int_{V} \frac{\left(\pi^{(\phi)}(p, \lambda)\right)^2}{ \left(s(\lambda)\right)^2} d^3p -  \frac{n_f}{\beta }- \left( S^x(\lambda)-S^0 \right), \text{and}
\end{aligned}
 \end{equation}

 \begin{equation} 
\begin{aligned}
\dot{s}(\lambda) := \left.\frac{d s(\lambda+ \epsilon)}{d \epsilon} \right|_{\epsilon=0} = \left.\frac{\partial S}{\partial \pi^{(s)}}\right|_{\lambda} = \frac{\pi^{(s)}(\lambda)}{m_s}.
\end{aligned}
\end{equation}
\normalsize

Given some state of the system, $\gamma(\lambda_0) \in \Gamma$  at $\lambda=\lambda_0$, the set of differential equations defined by equations (2) to (5) specifies the flow of the system from $\gamma(\lambda_0)$ to a new state $\gamma(\lambda_1) \in \Gamma$  at $\lambda=\lambda_1$. The flow generated by these differential equations can be integrated to generate a system trajectory $\gamma(\lambda) \in \Gamma$.

\section*{Observables and expectation values}
Observables for the system are functionals of the matter field, i.e. $\mathcal O \left[\phi \right] \in \mathbb{C}$. The expectation value of an observable can be calculated by averaging the observable along the trajectory of the system using the following equation:

 \begin{equation} 
\begin{aligned}
\left<\mathcal O \left[\phi\right]\right>_{\lambda}:= \lim_{\Lambda \to \infty}\frac{1}{\Lambda} \int_0^{\Lambda} \mathcal O \left[\phi(\lambda)\right] d \lambda.
\end{aligned}
\end{equation}

Assuming that the flow with respect to $\lambda$ is ergodic, one can also calculate expectation values of the observables using an ensemble average.  Because the action $S:= \left( s (\lambda)\left(S^x(\lambda)-S^0 \right)\right)$ is conserved and, by the definition of $S^0$, is equal to zero for all $\lambda$, the ensemble average is calculated by integrating over the variational phase space using the microcanonical ensemble probability measure defined by:

\begin{equation} 
\begin{aligned}
d\Gamma :=  C \, \delta \left( s \left(S^x-S^0 \right)\right) ds\,d\pi^{(s)}\,\mathcal{D}\left[\pi^{(\phi)}\right]\,\mathcal{D}\left[\phi \right]. 
\end{aligned}
\end{equation}

\noindent Here, $\mathcal{D}\left[\phi\right]:=\prod_{p \in V} d\phi(p)\}$ is the functional integral measure for the matter field $\phi$, $\mathcal{D}\left[\pi^{(\phi)}\right]$ is defined analogously, and $C$ is a constant. It can be shown that the partition function defined by $\mathcal{Z} := \int d \Gamma$ is given by

\begin{equation}
\begin{aligned} 
 \mathcal{Z} =Z_0 \int e^ {- \beta S^m[\phi] }\,\mathcal{D}\left[\phi \right] \text { where $Z_0$ is a constant.}
\end{aligned}
\end{equation}

To arrive at equation (8) from the definition $\mathcal{Z} := \int d \Gamma$, first make a change of variables $\underline{\pi}^{(\phi)} := \frac{\pi^{(\phi)}(x)}{s}$. This change of variables is equivalent to a re-scaling of $\lambda$. See, e.g., \cite{bond1999nose} for an analogous construction used in classical statistical mechanics. The change of variables changes the integration measure in the partition function by a factor of $s^{n_f}$ yielding:

\begin{equation}
\begin{aligned} 
\mathcal{Z} & := \int d \Gamma
\\& =C \int s^{n_f} \delta \left( s \left(S^x-S^0 \right)\right) ds\,d\pi^{(s)}\,\mathcal{D}\left[\underline{\pi}^{(\phi)}\right]\mathcal{D}\left[\phi \right].
\end{aligned}
\end{equation}

Next, integration over the Dirac delta function with respect to $ds$ can be performed using the identity $ \frac{ d }{ds} \delta\left[f(s)\right]= \frac{\delta\left[ f(s-s')\right]}{\left( \frac{df}{ds}|_{(s')}\right)}$, where $s'$ is the isolated zero of $f(s)$ given by the following equation:

\begin{equation}
\begin{aligned} 
s' =  exp  \bigg[  -\frac{\beta}{n_f} \bigg(\int_{V} \frac{1}{2} \left( \underline{\pi}^{(\phi)}(p)\right)^2 d^3p + \frac{\left(\pi^{(s)}\right)^2}{2m_s} +S^m[\phi]-S^0\bigg)  \bigg].
\end{aligned}
\end{equation}

Performing this integration with respect to $ds$ gives the following expression for $\mathcal{Z}$ where $C'$ is a new constant:
\small
\begin{equation}
\begin{aligned} 
 C' \int exp  \left[ - \beta \left( \int_{V} \frac{1}{2} \left( \underline{\pi}^{(\phi)}(p) \right)^2 d^3p + \frac{\left(\pi^{(s)}\right)^2}{2m_s}+S^m[\phi]   \right) \right]  d\pi^{(s)}\,\mathcal{D}\left[\underline{\pi}^{(\phi)}\right]\,\mathcal{D}\left[\phi \right].
\end{aligned}
\end{equation}
\normalsize

Because the integrals with respect to $d\pi^{(s)}$ and $\mathcal{D}\left[\underline{\pi}^{(\phi)}\right]$ are Gaussian, they can be evaluated to provide the desired equation for the partition function: 

\begin{equation}
\begin{aligned} 
 \mathcal{Z} =Z_0 \int e^{  - \beta S^m[\phi] }\,\mathcal{D}\left[\phi \right].
\end{aligned}
\end{equation}

The partition function is helpful for calculating expectation values because, by the assumption of ergodicity, the expectation value of some observable $\mathcal{O} \left[\phi \right] $ taken with respect to $\lambda$ agrees with its ensemble average. In equations:
\begin{equation} 
\left<\mathcal O \left[\phi\right]\right>_{\lambda}:= \lim_{\Lambda \to \infty}\frac{1}{\Lambda} \int_0^{\Lambda} \mathcal O \left[\phi(\lambda)\right] d \lambda,
\end{equation}

\begin{equation}
\begin{aligned} 
\left<\mathcal O \left[\phi \right]\right>_{\Gamma}:=
 Z_0 \int \mathcal O \left[\phi \right] e^{ - \beta  S^m[\phi] }\mathcal{D}\left[\phi \right], \text{and}
\end{aligned}
\end{equation}

\begin{equation}
\begin{aligned} 
 \left<\mathcal O \left[\phi \right]\right>_{\lambda}= \left<\mathcal O \left[\phi \right]\right>_{\Gamma}.
\end{aligned}
\end{equation}

\noindent Because the expectation value with respect to $\lambda$ agrees with the ensemble average, let $\left<\mathcal O \left[\phi \right]\right>$ denote the expectation value determined by either way. 

\section*{A bosonic Fock space of observables}

A bosonic Fock space can be constructed using the set of observables together with an inner product evaluated by taking the expectation value. To see this, define the inner product of observables by $\left<\mathcal O_1 ,\mathcal O_2 \right> := \left<\overline{\mathcal O_1 \left[\phi \right]}\mathcal O_2 \left[\phi \right] \right>$. This inner product has the following properties:

\begin{equation}
\begin{aligned} 
\left<\mathcal O_1,\mathcal O_2 \right>=\overline{\left<\mathcal O_2 ,\mathcal O_1 \right>},
\end{aligned}
\end{equation}

\begin{equation}
\begin{aligned} 
 \left<z \mathcal O_1 ,\mathcal O_2 \right>=\overline{z}\left<\mathcal O_1 ,\mathcal O_2  \right> \text { where } z \in \mathbb{C}, 
 \end{aligned}
\end{equation}

 \begin{equation}
\begin{aligned} 
\left<\mathcal O_1 ,z \mathcal O_2  \right>=z\left<\mathcal O_1 ,\mathcal O_2 \right> \text { where, again, } z \in \mathbb{C},
\end{aligned}
\end{equation}

\begin{equation}
\begin{aligned}  
\left<\mathcal O_1 +\mathcal O_2 , \mathcal O_3  \right>= \left<\mathcal O_1, \mathcal O_3  \right>+\left<\mathcal\mathcal O_2 , \mathcal O_3 \right>, 
 \end{aligned}
\end{equation}

 \begin{equation}
\begin{aligned} 
 \left<\mathcal O_1 , \mathcal O_2 + \mathcal O_3  \right>= \left<\mathcal O_1 , \mathcal O_2\right>+ \left<\mathcal O_1 , \mathcal O_3 \right>, 
\text{and}
\end{aligned}
\end{equation}

\begin{equation}
\begin{aligned} 
\left<\mathcal O_1 ,\mathcal O_1  \right> \geq 0.
\end{aligned}
\end{equation}

Thus, one can define a Hilbert space, $\mathcal{H}\left(\mathcal{M}\right)$, where the elements of the Hilbert space are the observables modulo those observables such that $\left<\mathcal O,\mathcal O \right> = 0$. 

The bosonic Fock space, denoted $\mathcal{F}\left(\mathcal{M}\right)$, can be constructed from the Hilbert space $\mathcal{H}\left(\mathcal{M}\right)$ using the construction provided in \cite{bar2007wave}. Briefly, the inner product on $\mathcal{H}$ induces an inner product, denoted $\left<F^{(j)}\big{|}  G^{(j)}  \right >$,  on elements, $F^{(j)}, G^{(j)}$, of the symmetric tensor product space, $\odot^j \mathcal{H}\left(\mathcal{M}\right)$, where $j$ is a positive integer. The inner product is defined by the following equations:

\begin{equation}
\begin{aligned}
\text{for  j $\geq$ 1:} & \left<F_1^{(1)} \odot . . . \odot F_j^{(1)} \big{|} G_1^{(1)}\odot . . . \odot G_j^{(1)}\right > 
\\& := \sum_{Perm\{1, . . .,j\}} \left<F_1^{(1)}, G_{Perm(1)}^{(1)}\right > . . .  \left<F_j^{(1)}, G_{Perm(j)}^{(1)}\right >,\text{ and}
\end{aligned}
\end{equation}

\begin{equation}
\begin{aligned} 
 &\text{ for j = 0:} \left<F^{(0)} | G^{(0)}\right > := \overline{F^{(0)}} G^{(0)} \text{ where }  F^{(0)}, G^{(0)} \in \odot^0 \mathcal{H}\left(\mathcal{M}\right) := \mathbb{C}.
\end{aligned}
\end{equation}

The bosonic Fock space is defined as the direct sum $\mathcal{F}\left(\mathcal{M}\right):=\oplus_{j=0}^{\infty} \odot^j \mathcal{H}\left(\mathcal{M}\right)$ and is a Hilbert space with the induced inner product, $\left< F | G \right >:= \sum_{j=0}^{\infty} \left<F^{(j)} |  G^{(j)}\right >$, where the elements, $F, G \in \mathcal{F}\left(\mathcal{M}\right)$, are sequences, $\left( F^{(0)}, F^{(1)}, F^{(2)}, . . .\right)$ and  $\left(G^{(0)}, G^{(1)}, G^{(2)}, . . .\right)$, with each $F^{(j)}, G^{(j)} \in  \odot^j \mathcal{H}\left(\mathcal{M}\right)$. The vacuum vector is given by $\Omega := \left(1, 0, 0, . . . \right)$. 

One can define creation and annihilation operators, $\hat{a}^*(\mathcal O)$ and $\hat{a}(\mathcal O)$ respectively, where $\mathcal O\in \mathcal{H}\left(\mathcal{M}\right)$. The action of $\hat{a}^*(\mathcal O)$ and $\hat{a}(\mathcal O)$ on each $\odot^j \mathcal{H}\left(\mathcal{M}\right) \subset \mathcal{F}\left(\mathcal{M}\right)$ is defined by the equations below where the tilde denotes that the term $\tilde{F }_i^{(1)}$ is excluded.

\begin{equation}
\begin{aligned} 
 \hat{a}^*(\mathcal O) : \left( F_1^{(1)} \odot . . . \odot F_j^{(1)} \right) \in \odot^j \mathcal{H} \left(\mathcal{M}\right)\rightarrow \left( \mathcal O \odot F_1^{(1)} \odot F_2^{(1)} . . . \odot F_j^{(1)}\right) \in \odot^{j+1} \mathcal{H}\left(\mathcal{M}\right)
\end{aligned}
\end{equation}

\small
\begin{equation}
\begin{aligned} 
\hat{a}(\mathcal O) : \left( F_1^{(1)} \odot . . . \odot F_j^{(1)} \right)  \in \odot^j \mathcal{H} \left(\mathcal{M}\right)\rightarrow \left(\sum_{i=1}^{j} \left<\mathcal O, F_i^{(1)}  \right>F_1^{(1)}  \odot . . . \odot \tilde{F }_i^{(1)} . . . \odot F_j^{(1)}  \right) \in \odot^{j-1} \mathcal{H}\left(\mathcal{M}\right)
\end{aligned}
\end{equation}
\normalsize

\begin{equation}
\begin{aligned} 
 \hat{a}(\mathcal O)\Omega :=0
\end{aligned}
\end{equation}

The creation and annihilation operators are adjoints of each other because $\left< \hat{a}^*(\mathcal O) F | G \right > = \left< F | \hat{a}(\mathcal O)G \right >$. See, e.g., \cite{bar2007wave} at equation 4.9. The creation and annihilation operators satisfy the canonical comutation relations $\left [\hat{a}(\mathcal O_1),\hat{a}(\mathcal O_2) \right]=0$, $\left [\hat{a}^*(\mathcal O_1),\hat{a}^*(\mathcal O_2) \right]=0$, and $\left [\hat{a}(\mathcal O_1),\hat{a}^*(\mathcal O_2) \right]=\left< \mathcal O_1, \mathcal O_2 \right>\hat{\mathcal{I}}_{\mathcal{F}\left(\mathcal{M}\right)}$, where $ \hat{\mathcal{I}}_{\mathcal{F}\left(\mathcal{M}\right)}$ is the identity operator on the Fock space $\mathcal{F}\left(\mathcal{M}\right)$. See, e.g., \cite{bar2007wave} at lemma 4.6.6.

\section*{``Smeared'' field observables and 2-point correlation functions}
A particularly interesting class of observables is the ``smeared'' field observables of the form $\phi(J) := \int \phi(x) J(x) d^4x$ where $\phi(x):= \frac{1}{8 \pi^3} \int_{\mathcal{\tilde{M}}} \phi(p) \delta_V e^{- i p \cdot x} d^4p$ is the Fourier transform of $\phi(p)$ with $\delta_V$ denoting the Dirac measure supported on $V := \{ p \in \mathcal{\tilde{M}} \, | \, P(p, \phi)=0 \}$ and $J \in \mathcal{C}^{\infty}_c \left(\mathcal{M}, \mathbb{C} \right)$ is a smooth, complex-valued function with compact support on the Minkowski spacetime $\mathcal{M}$. Note, in particular, that $\phi(x)$ satisfies the pseudodifferential equation $\tilde{P}(x)\phi(x)=0$ where $\tilde{P}(x) := \frac{1}{8 \pi^3} \int_{\mathcal{\tilde{M}}} P(p, \phi) e^{- i p \cdot x} d^4p$ because $\phi(p)$ is, by definition, not supported anywhere that $P(p, \phi) \neq 0$. For example, in the particular case of the real-valued matter field, $\phi(p)$, defined on $V_m^{free} := \{ p \in \mathcal{\tilde{M}} \, | \, p \cdot p -m^2=0  \text{ with } p_0 \geq 0\}$, the spacetime representation of the matter field, $\phi(x)$, satisfies the Klein-Gordon equation $\left( \Box + m^2 \right) \phi(x)=0$.

The inner product of two smeared field observables, $\left< \phi(J), \phi(K) \right>$, is equal to $\int \int \bar{J}(y) \left< \bar{\phi}(y)\phi(x)\right>K(x) \, d^4y \, d^4x$ where $ \left< \bar{\phi}(y)\phi(x)\right>$ is the 2-point correlation function of the statistical field theory. The 2-point correlation function $ \left< \bar{\phi}(y)\phi(x)\right>$  is just the spacetime representation of the momentum space correlation function $ \left< \phi(p')\phi(p)\right>$ and can be related to each other by $\frac{1}{64 \pi^6} \int  e^{ i p' \cdot y - i p \cdot x} \left< \bar{\phi}(p')\phi(p)\right> d^4p' d^4p$. The momentum space correlation function can be calculated by 

\begin{equation}
\begin{aligned} 
\left.\frac{\delta^2 ln \left( \mathcal{Z}\left[J\right]\right)}{\delta j(p') \delta j(p)}\right|_{j=0} =\left<\phi(p')\phi(p)\right>
\end{aligned}
\end{equation}

\noindent where $\mathcal{Z}\left[j \right]:= \left<e^{\int_V \phi(p) j(p) d^3p } \right>$ is the moment generating function with $j(p)$ a real-valued source field on $V$.

For the particular free field theory defined above on $V_m^{free} := \{ p \in \mathcal{\tilde{M}} \, | \, p \cdot p -m^2=0  \text{ with } p_0 \geq 0\}$ with $S^m_{free}[\phi]:=  \int_{V^{free}_m} \frac{1}{2} \left(\phi(p)\right)^2 d^3p$, the 2-point correlation function can be calculated exactly, assuming ergodicity \footnote{As discussed in the examples section, the free field system defined in this way may not be ergodic but a modified free field system still defined on $V_m^{fee}$ but with a matter action defined by $S^m_{free}[\phi]:=  \int_{V^{free}_m} \frac{1}{2} \left(\phi(p)\right)^2 d^3p+\frac{1}{2} \left(\int_{V^{free}_m} \phi(p)d^3p\right)^2$ may be.}. The  moment generating function of this free theory is equal to

\begin{equation}
\begin{aligned} 
 \mathcal{Z}^{free}\left[j \right] & \propto  \int e^{\int_{V^{free}_m} \phi(p) j(p) -   \frac{\beta}{2} \left( \phi(p)\right)^2 d^3p} \,\mathcal{D}\left[\phi \right].
 \end{aligned}
\end{equation}

\noindent The integral with respect to $\mathcal{D}\left[\phi \right]$ can be evaluated exactly to give

\begin{equation}
\begin{aligned} 
 \mathcal{Z}^{free}\left[j \right] & \propto  e^{\frac{1}{2 \beta}  \int_{V^{free}_m}  \left(j(p)\right)^2 d^3p}.
 \end{aligned}
\end{equation}

\noindent  Thus, the momentum space correlation function of this free theory is given by

\begin{equation}
\begin{aligned} 
\left.\frac{\partial^2 ln \left( \mathcal{Z}^{free}\left[j \right]\right)}{\partial j(p') \partial j(p)}\right|_{j=0} &=\left<\phi(p')\phi(p)\right>^{free}
\\ &=\frac{1}{2 \beta} \delta(p'-p) \text { on } V_m^{free}
\\& = \frac{1}{2 \beta} \delta_{V^{free}_m} \delta(p'-p)  \text { on } \mathcal{\tilde{M}}
\\& = \frac{1}{2 \beta} \theta(p_0) \delta(p \cdot p -m^2) \delta(p'-p) \text { by definition of }  \delta_{V^{free}_m}.
\end{aligned}
\end{equation}

Here,  $\theta$ is the Heaviside step function. Taking the Fourier transform of $\left<\phi(p')\phi(p)\right>^{free}$ gives the spacetime representation of the 2-point correlation function as $\left< \bar{\phi}(y)\phi(x)\right>^{free} = \frac{1}{2 \beta}  \Delta^+_m (y-x)$ where $\Delta^+_m (y-x):=\frac{1}{(2\pi)^3} \int \theta(p^0) \delta(p \cdot p -m^2) e^{-ip \cdot (y-x)} d^4p$ is the positive-frequency Wightman distribution. See, e.g., \cite{dutsch2019classical}.

\section*{Field operators}

Field operators are defined by $\hat{\phi}(J) :=  \hat{a}^*\left(\phi(J)\right)+\hat{a}\left(\phi(J)\right)$. These field operators can be considered to be the ``quantization'' of the smeared field observables $\phi(J)$ because they have many properties that one would like a quantum field theory to satisfy. 

For example, the field operators are self-adjoint because the creation and annihilation operators are adjoints of each other. See, e.g., \cite{bar2007wave}. Also, the field operators do not in general commute. Rather, the commutator of two field operators is given by $\left[\hat{\phi}(J), \hat{\phi}(K) \right]  = 2 i Im \left \{ \left< \phi(J), \phi(K) \right > \right \}  \hat{\mathcal{I}}_{\mathcal{F}}$. Moreover, if one defines Hermitian conjugation by $\hat{\phi}(J)^{\dag} :=\hat{\phi}(\bar{J})^*$, the field operators are Hermitian when J is real. As such, the field operators become Hermitian when restricting to the set of observables of the form $\phi(f) := \int \phi(x) f(x) d^4x$ where $f \in \mathcal{C}^{\infty}_c \left(\mathcal{M}, \mathbb{R} \right)$. With this restriction, the commutation relation between the field operators can be simplified to give $\left[\hat{\phi}(f), \hat{\phi}(g) \right]  = 2 i \int \int f(y) g(x) Im \left \{ \left< \bar{\phi}(y)\phi(x) \right>\right \} \, d^4y \, d^4x \; \hat{\mathcal{I}}_{\mathcal{F}\left(\mathcal{M}\right)}$. 

For the free field theory defined above, $Im \left \{ \left< \bar{\phi}(y)\phi(x) \right>^{free}\right \} = \frac{1}{2 \beta}  Im \left \{  \Delta^+_m (y-x)\right \}$ which is equal to $\frac{1}{2 \beta} $ times the Pauli-Jordan commutator function $\Delta_m(y-x):=\frac{-i}{(2 \pi)^3} \int sgn(p^0) \delta(p \cdot p-m^2) e^{-ip \cdot (y-x)} d^4p$. See, e.g., \cite{dutsch2019classical}. Because the Pauli-Jordan commutator function is equal to zero outside of the light cone, two field operators $\hat{\phi}(f)$ and $\hat{\phi}(g)$ commute when the supports of $f$ and $g$ are not connected by a causal path in $\mathcal{M}$. Thus, the free theory satisfies the microcausality condition. 

 \section*{The algebra of observables over a Cauchy Hypersurface}

Typically, in quantum field theory one works with an algebra of observables defined in terms of field operators over a Cauchy hypersurface instead of the algebra of observables defined in terms of operators on spacetime. One can show, at least in the case of the free theory discussed above, that there is an isomorphism of the algebra of field operators generated by the smeared field observables of the form $\phi(f) := \int \phi(x) f(x) d^4x$ where $f \in \mathcal{C}^{\infty}_c \left(\mathcal{M}, \mathbb{R} \right)$ modulo the nullspace of $\left<\phi(f),\phi(g)\right>^{free}$, and an algebra of field operators over a Cauchy hypersurface $\Sigma_t$.

To see this, first define the half-spaces $\mathcal{M}^+ :=\{x \in \mathcal{M} \, | \, x_0 \geq t \}$ and $\mathcal{M}^- :=\{x \in \mathcal{M} \, | \, x_0 \leq t \}$. The boundary of each of these half-spaces  is the Cauchy hypersurface $\Sigma_t$.  Next, define  $u(y) := \int \left< \bar{\phi}(y)\phi(x)\right>^{free}  g(x) \, d^4x$ so that $\left<\phi(f),\phi(g)\right>^{free}$ is equal to $\int f(y) u(y)  \, d^4y$.

Then, define $w^{+}(y) := \int_{\mathcal{M}^+} \Delta^{ret}_m(y-x) f(x)  \, d^4x$ and $w^{-}(y) := \int_{\mathcal{M}^-} \Delta^{adv}_m(y-x) f(x)  \, d^4x$ where $\Delta^{adv}_m(y-x)$ and $\Delta^{ret}_m(y-x)$ are the advanced and retarded Green's functions for the Klein-Gordon equation. See, e.g., \cite{dutsch2019classical} and \cite{bar2007wave}. Proceeding similarly to lemma 3.2.2 in \cite{bar2007wave}, 

\begin{equation}
\begin{aligned} 
\int \left( \left(\Box +m^2\right) \left(w^+(y)+w^-(y)\right) \right) u(y) &- \left(w^+(y)+w^-(y)\right) \left( \left(\Box +m^2\right) u(y) \right) \, d^4y 
\\ &=\int f(y) u(y)\, d^4y 
\\& = \int \int f(y)  \left< \bar{\phi}(y)\phi(x)\right>^{free}  g(x)  \, d^4y \, d^4x 
\\& = \left<\phi(f),\phi(g)\right>^{free} \text { , and}
\end{aligned}
\end{equation}

\begin{equation}
\begin{aligned} 
& \int_{\mathcal{M}^{\pm}} \left( \left(\Box +m^2\right) w^{\pm}(y) \right) u(y) - w^{\pm}(y) \left( \left(\Box +m^2\right) u(y) \right) \, d^4y 
\\ &=-\int_{\mathcal{M}^{\pm}}  \partial^{\mu} \left( \left( \partial_{\mu} w^{\pm}(y) \right) u(y) - w^{\pm}(y) \left( \partial_{\mu} u(y) \right) \right) \, d^4y 
\\& = \mp \int_{\Sigma_t} \left(\partial_{0} w^{\pm}(y) \right) u(y) - w^{\pm}(y) \left( \partial_{0} u(y) \right) \, d^3y.
\end{aligned}
\end{equation}

\noindent Thus,

\begin{equation}
\begin{aligned} 
\left<\phi(f),\phi(g)\right>^{free} & =  \int_{\Sigma_t} \left(\partial_{0} w^{-}(y) \right) u(y) - w^{-}(y) \left( \partial_{0} u(y) \right) \, d^3y
\\& \;  - \int_{\Sigma_t} \left(\partial_{0} w^{+}(y) \right) u(y) - w^{+}(y) \left( \partial_{0} u(y) \right) \, d^3y
\\& = \int_{\Sigma_t} \partial_{0} \left( \int \Delta_m(y-x) f(x )d^4x \right) u(y) 
\\& \; - \left( \int \Delta_m(y-x) f(x )d^4x \right) \left( \partial_{0} u(y) \right) \, d^3y.
\end{aligned}
\end{equation}

Given some Cauchy data, $\left( v_0 , v_1 \right)$, having compact support on $\Sigma_t$, there is a unique $f(x) \in \frac{C^{\infty}_c\left( \mathcal{M}, \mathbb{R}\right) }{\left( \Box+m^2 \right)C^{\infty}_c\left( \mathcal{M}, \mathbb{R}\right)}$ satisfying $v_0 := \left. \left( \int \Delta_m(y-x) f(x )d^4x \right) \right|_{\Sigma_t}$ and $v_1 := \left. \left( \partial_0 \int \Delta_m(y-x) f(x) d^4x \right) \right|_{\Sigma_t}$. See, e.g., \cite{bar2007wave} at theorem 3.2.11. This defines a map $SOLVE: \left( v_0 , v_1 \right) \to f(x)$. This map can then be composed with another map $W: f(x) \to \left( v'_0 , v'_1 \right) $ defined by  $v'_0(y) = \left. \left(\int \left< \bar{\phi}(y)\phi(x)\right>^{free}  f(x) d^4x \right)\right|_{\Sigma_t}$ and $v'_1(y) = \left. \left(\partial_0 \int \left< \bar{\phi}(y)\phi(x)\right>^{free}  f(x) d^4x \right)\right|_{\Sigma_t}$. The composition of these two maps sends Cauchy data with compact support on $\Sigma_t$ to a pair of functions on $\Sigma_t$ that do not in general have compact support.  Nonetheless, given any two sets of Cauchy data, $\left( v_0 , v_1 \right)$ and  $\left( u_0 , u_1 \right)$, having compact support on $\Sigma_t$, one can define an inner product by

\begin{equation}
\begin{aligned} 
& \left<\left( v_0 , v_1 \right) , \left( u_0 , u_1 \right)\right>^{free}_{\Sigma_t}  := \int_{\Sigma_t} v_1(y) u'_0(y) - v_0(y) u'_1(y) \, d^3y \text { where }
\\ & \left( u'_0 , u'_1 \right) := W \circ SOLVE \circ  \left( u_0 , u_1 \right).
\end{aligned}
\end{equation}

By these definitions then 

\begin{equation}
\begin{aligned} 
& \left<\left( v_0 , v_1 \right) , \left( u_0 , u_1 \right)\right>^{free}_{\Sigma_t}  = \left<\phi(f),\phi(g)\right>^{free} \text { where}
\\& f(x) = SOLVE \circ  \left( v_0 , v_1 \right) \text{, and}
\\& g(x) = SOLVE \circ  \left( u_0 , u_1 \right).
\end{aligned}
\end{equation}

Thus, the set of all Cauchy data, $\left( v_0 , v_1 \right)$, having compact support on $\Sigma_t$, modulo those lying in $Ker \left( W \circ SOLVE \right)$, equipped with the inner product $\left<\left( v_0 , v_1 \right) , \left( u_0 , u_1 \right)\right>^{free}_{\Sigma_t} $ defines a Hilbert space, $\mathcal{H}\left( \Sigma_t \right)$. This Hilbert space $\mathcal{H}\left( \Sigma_t \right)$ is isometric to the Hilbert space $\mathcal{H}\left( \mathcal{M} \right)$, with inner product  $\left<\phi(f),\phi(g)\right>^{free}$, consisting of functions $f,g \in C^{\infty}_c\left( \mathcal{M}, \mathbb{R}\right)$ modulo the nullspace of $\left<\phi(f),\phi(g)\right>^{free}$. Accordingly, the algebra of observables defined in terms of $\mathcal{H}\left( \Sigma_t \right)$ and $\mathcal{H}\left( \mathcal{M}\right)$ are isomorphic. 

Because this construction did not depend on the choice of Cauchy hypersurface, one could replicate the same construction for another Cauchy hypersurface, such as  $\Sigma_{t'}$, to given another Hilbert space, $\mathcal{H}\left( \Sigma_{t'} \right)$, isometric with the first one, $\mathcal{H}\left( \Sigma_t \right)$. There thus exist unitary time translation operators, $U(t',t))$, mapping elements of  $\mathcal{H}\left( \Sigma_t \right)$ to elements of $\mathcal{H}\left( \Sigma_{t'} \right)$. 

\section*{Examples}

An advantage of the relativistic statistical field theory developed in the previous sections is that it is easy to discretize and simulate numerically. In each of the following examples, $V$ is a regular 3-dimensional Euclidean lattice of $N=25 \times 25 \times 25$ equally spaced points $p:=(p_1,p_2,p_3) \in V$ having a lattice spacing of $\pm 0.1$. The number of degrees of freedom, $n_f$, for these examples is equal to N (i.e. 1 real degree of freedom per point in the discretized V). The real-valued scalar field $\phi$ was initialized with a value of 0 at each point. The variational conjugate field $\pi^{(\phi)}$ was initialized at each point with a value randomly selected from the interval $[-2.5, 2.5]$ so that $\frac{\sum \pi^{(\phi)}}{2 N}$ was about equal to 1. For every example, $s$ was initialized at 1,  $\pi^{(s)}$ was initialized at 0, $\beta$ was set equal to 1, the mass term $m$ was equal to 1, and $m_s$ was set equal to $N$. 

Once an initial configuration was established, each example was numerically integrated with respect to $\lambda$ using the explicit leap-frog algorithm provided by Bond, Leimkuhler, and Laird \cite{bond1999nose} with a step size of $\Delta \lambda=0.01$. Each system was ``equilibrated" by stepping the system forward for 1,000,000 steps. After this ``equilibration'' period, each system was further numerically integrated with respect to $\lambda$ for an additional 1,000,000 steps during which data was collected.
 \pagebreak
\subsection*{\small \;\;Example 1:\;}

In this example, a free field theory was simulated using $V=V_m^{free} := \{ p \in \mathcal{\tilde{M}} \, | \, p \cdot p -m^2=0  \text{ with } p_0 \geq 0\}$ with $S^m_{free}[\phi]:=  \sum_{p \in V^{free}_m} \frac{1}{2} \left(\phi(p, \lambda)\right)^2$. The spacetime 2-point correlation function $\left< \bar{\phi}(y)\phi(0)\right>^{free}$ was then calculated as $\frac{1}{\Lambda} \sum_{\lambda=0}^{\Lambda} \sum_{p',p} \phi(p', \lambda) \phi(p, \lambda) exp\left(i \left( \sqrt{ \left(p_1^2+p_2^2+p_3^2+m^2\right)} \, y_0-p_1 \, y_1-p_2 \, y_2-p_3 \, y_3\right)\right)$.

Figure 1A depicts the real part of 2-point correlation function on a space-time plane. Figure 1B depicts the imaginary part of 2-point correlation function on the space-time plane. Figure 1C depicts a line cut of the 2-point correlation function in the space-time plane along the space axis at time equal to zero. Figure 1D depicts a line cut of the 2-point correlation function in the space-time plane along the time axis and at the space parameter equal to zero. As can be seen in figures 1A-1D, the 2-point correlation function does not look qualitatively like the expected positive frequency Wightman distribution. It is believed that this discrepancy is due to this system not being ergodic. That is, each of the different  modes $ \phi(p)$ in momentum space are independent of each other. Thus, there is no phase-space mixing during the evolution of the system with respect to the variational parameter $\lambda$. This hypothesis was tested in the next example.

\begin{figure*}[b!]
\centering
\includegraphics[width=0.75\textwidth]{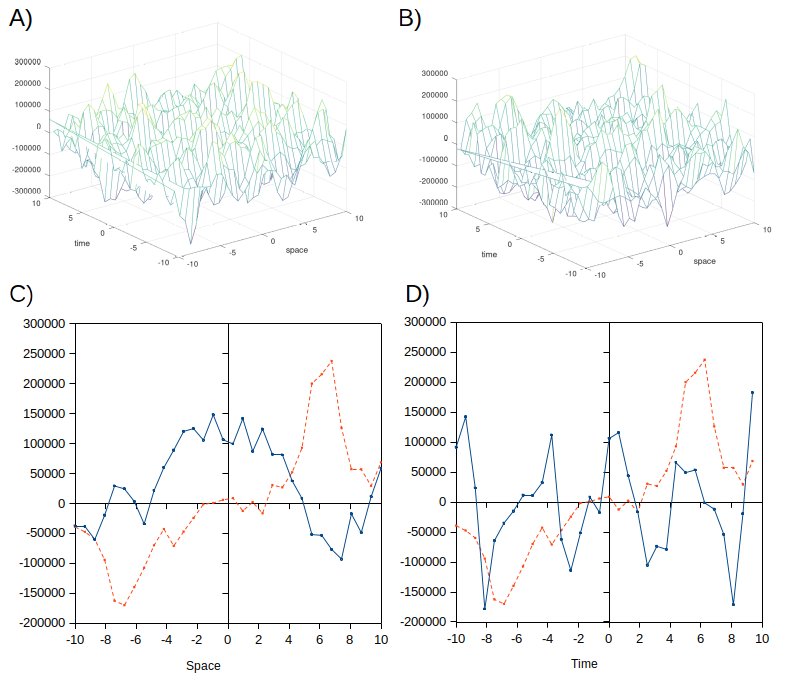}
\caption{A, B) real, imaginary parts of the 2-point correlation function in a space-time plane, respectively. C, D) space-like line cut, time-like line cut of the 2-point correlation function in the space-time plane, respectively, showing real (\includegraphics[scale=0.1]{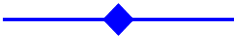}) and imaginary (\includegraphics[scale=0.1]{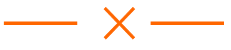}) parts.}
\end{figure*}
 \pagebreak

\subsection*{\small \;\;Example 2:\;}

The only difference between example 2 and example 1 is that example 2 used a different matter action defined by $S^m_{free}[\phi]:=  \sum_{p \in V^{free}_m} \frac{1}{2} \left(\phi(p, \lambda)\right)^2+ \frac{1}{2} \left(\sum_p\phi(p, \lambda)\right)^2$. The newly added term $\frac{1}{2} \left(\sum_p\phi(p, \lambda)\right)^2$ allows the different modes in momentum space to interact so that the system can evolve with more phase-space mixing. The expected 2-point correlation function in example 2 is still controlled by the embedding of $V$ in $\mathcal{M}$. Thus, the 2-point correlation function in example 2 is again expected to be proportional to the positive frequency Wightman distribution. 

Figure 2A depicts the real part of 2-point correlation function on a space-time plane. Figure 2B depicts the imaginary part of 2-point correlation function on the space-time plane. Figure 2C depicts a line cut of the 2-point correlation function in the space-time plane along the space axis at time equal to zero. Figure 2D depicts a line cut of the 2-point correlation function in the space-time plane along the time axis and at the space parameter equal to zero. As can be seen in figures 2A-2D, the 2-point correlation function does  look qualitatively like the expected positive frequency Wightman distribution. This suggests that the added term $\frac{1}{2} \left(\sum_p\phi(p, \lambda)\right)^2$ does improve mixing to afford a more ergodic averaging over the variational phase-space. 
\begin{figure*}[b!]
\centering
\includegraphics[width=0.65\textwidth]{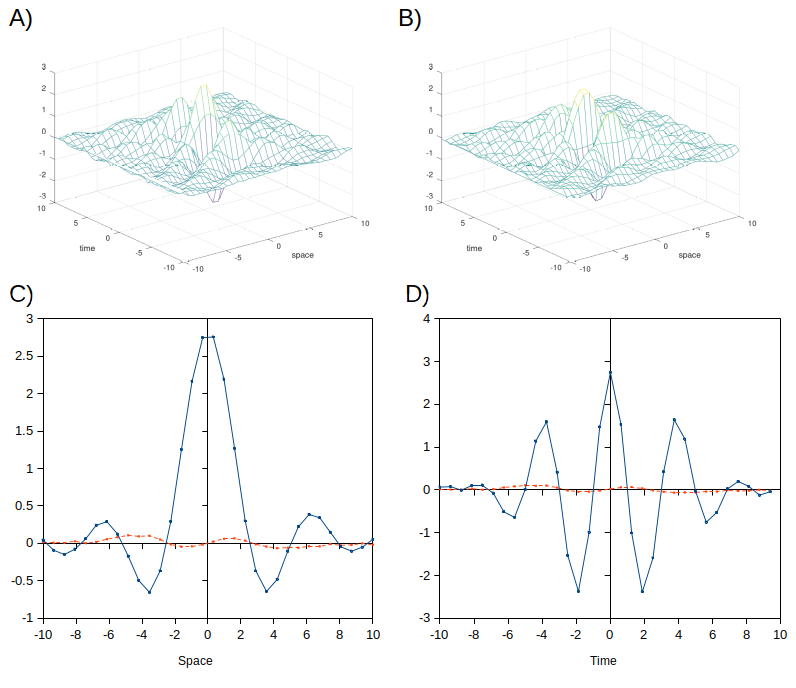}
\caption{A, B) real, imaginary parts of the 2-point correlation function in a space-time plane, respectively. C, D) space-like line cut, time-like line cut of the 2-point correlation function in the space-time plane, respectively, showing real (\includegraphics[scale=0.1]{Re}) and imaginary (\includegraphics[scale=0.1]{Im}) parts.}
\end{figure*}
 \pagebreak

\subsection*{\small \;\;Example 3:\;}

The only difference between example 3 and example 2 is that example 3 used  a dynamically defined $V := \{ p \in \mathcal{\tilde{M}} \, | \, p \cdot p -\left(\sum_p\phi(p, \lambda)\right)^2=0  \text{ with } p_0 \geq 0\}$. The spacetime 2-point correlation function $\left< \bar{\phi}(y)\phi(0)\right>$ was then calculated as 
\scriptsize
\begin{equation}
\begin{aligned} 
& \left< \bar{\phi}(y)\phi(0)\right> =
\\& \frac{1}{\Lambda} \sum_{\lambda=0}^{\Lambda} \sum_{p',p} \phi(p', \lambda) \phi(p, \lambda) exp\left(i \left( \sqrt{ \left(p_1^2+p_2^2+p_3^2+\left(\sum_p\phi(p, \lambda)\right)^2\right)} \, y_0-p_1 \, y_1-p_2 \, y_2-p_3 \, y_3\right)\right).
\end{aligned}
\end{equation}
\normalsize

\noindent In this way the mass term is a dynamic quantity determined by the system.

Figure 3A depicts the real part of 2-point correlation function on a space-time plane. Figure 3B depicts the imaginary part of 2-point correlation function on the space-time plane. Figure 3C depicts a line cut of the 2-point correlation function in the space-time plane along the space axis at time equal to zero. Figure 3D depicts a line cut of the 2-point correlation function in the space-time plane along the time axis and at the space parameter equal to zero. As can be seen in figures 3A-3D, the 2-point correlation function does  look qualitatively like a superposition of positive frequency Wightman distributions having different mass terms. This gives a nonlinear theory with a 2-point correlation function that is some weighted average of positive frequency Wightman distributions.
\begin{figure*}[b!]
\centering
\includegraphics[width=0.6\textwidth]{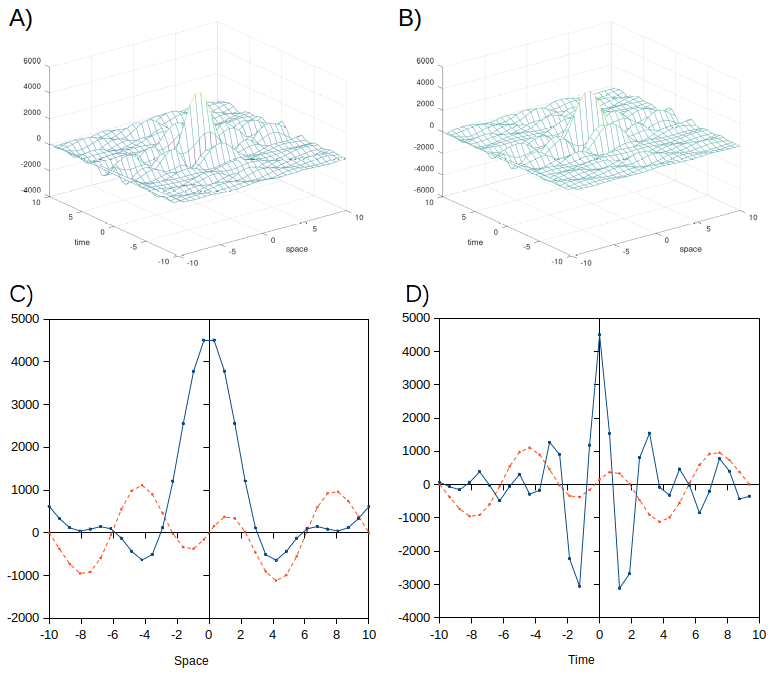}
\caption{A, B) real, imaginary parts of the 2-point correlation function in a space-time plane, respectively. C, D) space-like line cut, time-like line cut of the 2-point correlation function in the space-time plane, respectively, showing real (\includegraphics[scale=0.1]{Re}) and imaginary (\includegraphics[scale=0.1]{Im}) parts.}
\end{figure*}
 \pagebreak
\subsection*{\small \;\;Example 4:\;}

The only difference between example 4 and example 2 is that example 4 used used a locally defined dyanmical $V := \{ p \in \mathcal{\tilde{M}} \, | \, p \cdot p -\left(\phi(p, \lambda)\right)^2=0  \text{ with } p_0 \geq 0\}$. The spacetime 2-point correlation function $\left< \bar{\phi}(y)\phi(0)\right>$ was then calculated as 

\begin{equation}
\begin{aligned} 
& \left< \bar{\phi}(y)\phi(0)\right> =
\\& \frac{1}{\Lambda} \sum_{\lambda=0}^{\Lambda} \sum_{p',p} \phi(p', \lambda) \phi(p, \lambda) exp\left(i \left( \sqrt{ \left(p_1^2+p_2^2+p_3^2+\left(\phi(p, \lambda)\right)^2\right)} \, y_0-p_1 \, y_1-p_2 \, y_2-p_3 \, y_3\right)\right).
\end{aligned}
\end{equation}

\noindent In this way the mass term is a dynamic quantity determined locally for each point in $V$ by the system.

Figure 4A depicts the real part of 2-point correlation function on a space-time plane. Figure 4B depicts the imaginary part of 2-point correlation function on the space-time plane. Figure 4C depicts a line cut of the 2-point correlation function in the space-time plane along the space axis at time equal to zero. Figure 4D depicts a line cut of the 2-point correlation function in the space-time plane along the time axis and at the space parameter equal to zero. As can be seen in figures 4A-4D, the 2-point correlation function does  look qualitatively like a superposition of positive frequency Wightman distributions having different mass terms. Again, this gives a nonlinear theory with a 2-point correlation function that is some weighted average of positive frequency Wightman distributions. Here, in example 4, unlike in example 3, the weighted average of positive frequency Wightman distributions is determined locally on momentum space.
\begin{figure*}[b!]
\centering
\includegraphics[width=0.5\textwidth]{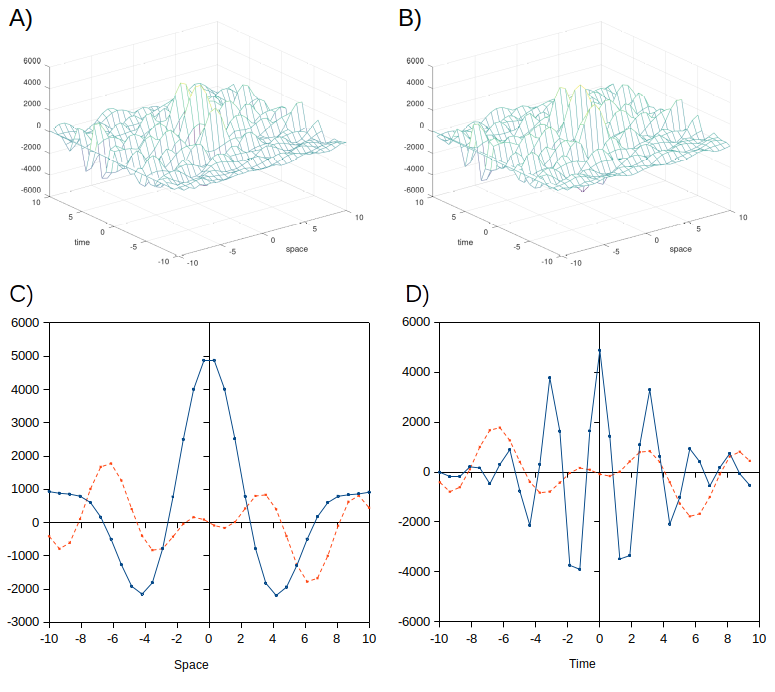}
\caption{A, B) real, imaginary parts of the 2-point correlation function in a space-time plane, respectively. C, D) space-like line cut, time-like line cut of the 2-point correlation function in the space-time plane, respectively, showing real (\includegraphics[scale=0.1]{Re}) and imaginary (\includegraphics[scale=0.1]{Im}) parts.}
\end{figure*}
 \pagebreak
\section*{Discussion}

This article demonstrates that a relativistic statistical field theory can be constructed for fluctuating fields on momentum space. The relativistic statistical field theory has an algebra of observables with properties, such as microcausality, that are similar to the types of properties one would expect of a quantum field theory. Numerical simulations of the fluctuating fields demonstrate proof-of-concept use of this relativistic statistical field theory formulation in non-perturbative calculations and the resulting 2-point correlation functions look qualitatively correct, at least for the free scalar field system. Further developments may, however, be needed to demonstrate whether a relativistic statistical field theory can be constructed to have all of the properties of quantum field theory and, thus, provide an equivalent reformulation.

\section*{Author Contact Information}
\noindent Brenden.McDearmon@gmail.com
\bibliographystyle{unsrt}
\bibliography{citations}

\end{document}